\documentclass[twocolumn,showpacs,prb,floatfix,citeautoscript,cite]{revtex4}
\usepackage{graphicx}
\usepackage{epstopdf}
\usepackage{gensymb}
\usepackage{color}
\usepackage{amsmath}

\begin{document}

\title{$Ab$ $initio$ methodology for magnetic exchange parameters: \\Generic four-state energy mapping on Heisenberg spin Hamiltonian}

\author{D. \v{S}abani}
\affiliation{Department of Physics, University of Antwerp, Groenenborgerlaan
171, B-2020 Antwerp, Belgium}
\affiliation{NANOlab Center of Excellence, University of Antwerp, Belgium}

\author{C. Bacaksiz}
\affiliation{Department of Physics, University of Antwerp, Groenenborgerlaan
171, B-2020 Antwerp, Belgium}
\affiliation{NANOlab Center of Excellence, University of Antwerp, Belgium}

\author{M. V. Milo\v{s}evi\'c}
\email{milorad.milosevic@uantwerpen.be}
\affiliation{Department of Physics, University of Antwerp, Groenenborgerlaan
171, B-2020 Antwerp, Belgium}
\affiliation{NANOlab Center of Excellence, University of Antwerp, Belgium}

\pacs{75.10.Hk, 75.30.-m, 75.40.Mg, 71.15.Mb, 71.70.Gm, 11.30.Ly}

\date{\today}

\begin{abstract}
The recent development in the field of 2D magnetic materials urges reliable theoretical methodology for determination of magnetic properties. Among the available methods, $ab$ $initio$ four-state energy mapping based on Density Functional Theory stands out as a powerful technique to calculate the magnetic exchange interaction in the Heisenberg spin model. Although the required formulae were explained in earlier works, the considered Hamiltonian in those studies always corresponded to the specific case that $J$-matrix is anti-symmetric, which may be misleading in other cases. Therefore, using the most general form of the Heisenberg spin Hamiltonian, we here derive the generic formulae. With a proper choice of four different magnetic states, a single formula governs all elements of the exchange interaction matrix for any considered pair of spin sites.
\end{abstract}

\maketitle
\section{INTRODUCTION}\label{intro}

After the experimental realization of the ferromagnetic monolayer of 
CrI$_3$ in 2017,\cite{Huang_2Dmag} two-dimensional (2D) magnetic materials have been at the very forefront of both theoretical\cite{Gibertini_magn2D,Sivadas,Lado,Xu_Kitaev} and experimental\cite{Gong,Deng_gatetune,Bonilla_VSe} investigations. Previously, 2D ferromagnetism was not deemed possible due to Mermin-Wagner theorem,\cite{MWtheorem} which states that long-range order cannot survive temperature fluctuations in an isotropic system. In monolayer CrI$_3$ and in different 2D magnetic materials reported later,\cite{Bonilla_VSe,Gibertini_magn2D} the magnetic anisotropy due to the strong spin-orbit coupling removes the Mermin-Wagner restriction and allows the material to retain the magnetization at non-zero temperature. 

On the other hand, the strong spin-orbit coupling also instigates antisymmetric magnetic 
exchange interaction between the spin sites, the so-called Dzyaloshinskii-Moriya 
interaction (DMI).\cite{Moriya_DMI,Dzia_DMI} Contrary to usual magnetic 
exchange interaction which favors the magnetic moments to be parallel or 
anti-parallel, DMI forces the magnetic moments to be orthogonal. DMI is observed in the presence of structural anisotropy in the system at low temperatures and is responsible for emergence of non-trivial spin textures, such as skyrmionic ones.\cite{Skyrmions,Yang}

With such prominent recent advances in the field of 2D magnetism and the emerging 
significance of the microscopic interactions, such as DMI, on the overall 
magnetic properties, the need for reliable theoretical frameworks grows 
rapidly. The calculation of the microscopic magnetic parameters using methods based on Density Functional Theory (DFT) is already common.\cite{Sivadas,Chittari,Gong,Lado,E-mapping0,E-mapping} These methods, in general, rely on obtaining the energies of several alternative magnetic configurations and mapping those energies to the specific Hamiltonian that governs the considered system. Such energy mapping produces a system of equations to be solved by algebraic methods in order to obtain the magnetic exchange parameters. Among such methods, the four-state methodology (4SM), which was presented by Xiang \textit{et al.}\cite{E-mapping0,E-mapping}, stands out as particularly effective.\cite{Gong,Xu_Kitaev} One should note however that 4SM, though widely used in the analysis of 2D magnetic crystal analysis, is not limited to 2D crystals exclusively. As explained in Refs. ~\onlinecite{E-mapping0,E-mapping}, 4SM can be employed on any magnetic system. 

The advantage of the 4SM is that all parameters are calculated pair- or site-wise, instead of collective spin state considerations leading to quasi-averaged values of the magnetic exchange parameters. Furthermore, Refs. ~\onlinecite{E-mapping0,E-mapping} also presented the technique to calculate DMI parameters in detail. As a consequence of the recent advance in field of 2D-magnetism, the impact of studies involving magnetic chirality grows rapidly.\cite{SE1,SE2,Xu_Kitaev} However, in both Refs. ~\onlinecite{E-mapping0} and ~\onlinecite{E-mapping}, the considered spin Hamiltonian consisted of an exchange interaction matrix where the off-diagonal elements were antisymmetric. In Ref. ~\onlinecite{gen_notgenH} (equation (2c)) and Ref. ~\onlinecite{gen_notgenH1}, one can find that antisymmetric spin Hamiltonian is considered as a general one. This is not correct and in the later studies,\cite{Xu_Kitaev,SE1,SE2} such consideration was shown not to be compatible with all materials, although in a subtle way i.e. without any specific comment on that matter. 

To remove any possible ambiguity in the procedure to properly calculate off-diagonal exchange elements and DMI components, here we rewrite the methodology by considering the truly general Heisenberg spin Hamiltonian, without any constraint. In a general case, one should use the same formulae for all exchange parameters $J_{\alpha\beta}$ (without extra minus sign in any case), and then calculate DMI components in e.g. Cartesian frame of reference as $D_{x} = \frac{1}{2}(J_{yz}-J_{zy})$, 
$D_{y} = \frac{1}{2}(J_{zx}-J_{xz})$, and
$D_{z} = \frac{1}{2}(J_{xy}-J_{yx})$. In addition, we further derive the formulae for single-ion anisotropy (SIA) parameters. We also discuss formation of the SIA matrix in the presence of rotational symmetry as found in 2D and quasi-2D materials (such as few-layer 2D materials and layered bulk materials), where it is clearly possible to distinct the in-plane directions from out-of-plane direction. However, different constraints for SIA matrix might be imposed by different symmetry operations, even in 3D crystals, as for example considered by Xu \textit{et al.}\cite{SE1} for the rotation around (111) direction. Although 2D magnetism was the core motivation for this article, the formulae presented here are applicable to a magnetic system without \textit{any} structural constraint.

The paper is organized as follows. In Sec. \ref{method}, starting from the most general form of the spin Hamiltonian, we re-derive the formulae that allow one to calculate the exchange parameters between different spin sites as presented in Sec. \ref{driveJ}, as well as the single-ion anisotropy as presented in Sec. \ref{driveA}. In Sec. \ref{example} we present the application of the 4SM methodology on monolayer CrI$_3$, with matrix elements extracted from DFT. We present the results obtained using our derived formulae and the results obtained by blindly following the formulae from Refs. ~\onlinecite{E-mapping0,E-mapping}, and we illustrate in detail how one can get misled in the formalism and obtain erroneous conclusions about the physics of magnetic systems. Finally, our summary and conclusions are given in Sec. \ref{Conc}.

\section{Re-derivation of the formulae}\label{method}

In this chapter, we present the derivation of the formulae for exchange ($J$) and single-ion anisotropy (A)  parameters, starting from the general Heisenberg spin Hamiltonian for a magnetic system given as:  
\begin{eqnarray}\label{prob2}
H&=&H_{EX}+H_{SIA} \nonumber \\ \nonumber \\ 
&=&\sum_{i\textless j} \vec{S_{i}} \cdot J_{ij} \cdot \vec{S_{j}} + \sum_{i} \vec{S_{i}} \cdot A_{ii} \cdot \vec{S_{i}}.
\end{eqnarray}
The first term describes the magnetic exchange interaction between $i$th and $j$th spin sites with $J_{ij}$ being $3 \times 3$ matrix. In this study, spin is considered as a classical vector $\vec{S_{i}} = (S_{i}^{x}, S_{i}^{y}, S_{i}^{z})$ where $x$, $y$, and $z$ are chosen Cartesian coordinates. For instance, the interaction  between $S_{i}^{y}$ and $S_{j}^{z}$ is determined by $J_{ij}^{yz}$, or equivalently by $J_{ji}^{zy}(\equiv J_{ij}^{yz})$. 

The second term in the Hamiltonian describes the interaction between the spin components of a single ion. $A_{ii}$ is also a $3 \times 3$ matrix which consists of the elements $A_{ii}^{\alpha\beta}$ ($\alpha$ and $\beta$ are Cartesian coordinates).

\subsection{Exchange parameter (J)}\label{driveJ}

We start with the exchange term. In order to determine the magnetic exchange interaction between the two neighbouring spins, one needs to calculate all nine parameters (nine elements of exchange interaction matrix $J_{ij}$). The exchange part of the general Hamiltonian can be written in the explicit form:

\begin{eqnarray}\label{prob3}
H_{EX}&=&\sum_{i\textless j} \vec{S_{i}} \cdot J_{ij} \cdot \vec{S_{j}}\nonumber \\
&=&\sum_{i\textless j} \Big[ S_{i}^{x}\cdot J_{ij}^{xx}\cdot S_{j}^{x}
+ S_{i}^{x}\cdot J_{ij}^{xy}\cdot S_{j}^{y} 
+ S_{i}^{x}\cdot J_{ij}^{xz}\cdot S_{j}^{z}  
\nonumber \\ \nonumber \\ 
&+& S_{i}^{y}\cdot J_{ij}^{yx}\cdot S_{j}^{x}
+  S_{i}^{y}\cdot J_{ij}^{yy}\cdot S_{j}^{y} 
+  S_{i}^{y}\cdot J_{ij}^{yz}\cdot S_{j}^{z}  \nonumber \\ \nonumber \\ 
&+& S_{i}^{z}\cdot J_{ij}^{zx}\cdot S_{j}^{x}
+  S_{i}^{z}\cdot J_{ij}^{zy}\cdot S_{j}^{y}
+  S_{i}^{z}\cdot J_{ij}^{zz}\cdot S_{j}^{z}  \Big].
\end{eqnarray}

One should note that in general $J_{ij}^{\alpha\beta} \neq J_{ij}^{\beta\alpha}$ ($\alpha, \beta = x, y, z; \alpha \neq \beta$), though the symmetry of a system might impose equality. Next we arbitrary choose two spin sites (labeled as $i=1$ and $j=2$). The contribution of the chosen pair to the general Hamiltonian can be written as:

\begin{eqnarray}\label{prob6}
H &=& \vec{S_{1}}\cdot J_{12}\cdot \vec{S_{2}} 
+ \sum_{j \neq 2} \vec{S_{1}}\cdot J_{1j}\cdot \vec{S_{j}} 
+ \sum_{i \neq 1} \vec{S_{i}}\cdot J_{i2}\cdot \vec{S_{2}} \nonumber \\ 
&+& \sum_{i\neq 1,j\neq 2} \vec{S_{i}}\cdot J_{ij}\cdot \vec{S_{j}} 
+ \vec{S_{1}}\cdot A_{11}\cdot \vec{S_{1}} 
+ \vec{S_{2}}\cdot A_{22}\cdot \vec{S_{2}} \nonumber \\ 
&+& \sum_{i\neq 1,2} \vec{S_{i}}\cdot A_{ii}\cdot \vec{S_{i}}.
\end{eqnarray}
The decomposition of the matrices into Cartesian components results in:
\begin{eqnarray}\label{prob7}
H &=& \sum_{\alpha,\beta} S_{1}^{\alpha}\cdot J_{12}^{\alpha\beta}\cdot S_{2}^{\beta} + \sum_{j \neq 2}\sum_{\alpha,\beta} S_{1}^{\alpha}\cdot J_{1j}^{\alpha\beta}\cdot S_{j}^{\beta} \nonumber
\\
&+& \sum_{i \neq 1}\sum_{\alpha,\beta} S_{i}^{\alpha}\cdot J_{i2}^{\alpha\beta}\cdot S_{2}^{\beta} + \sum_{i \neq 1, j \neq 2}\sum_{\alpha,\beta} S_{i}^{\alpha}\cdot J_{ij}^{\alpha\beta}\cdot S_{j}^{\beta} \nonumber
\\
&+& \sum_{\alpha,\beta} S_{1}^{\alpha}\cdot A_{11}^{\alpha\beta}\cdot S_{1}^{\beta} + \sum_{\alpha,\beta} S_{2}^{\alpha}\cdot A_{22}^{\alpha\beta}\cdot S_{2}^{\beta} \nonumber
\\
&+& \sum_{i \neq 1,2}\sum_{\alpha,\beta} S_{i}^{\alpha}\cdot A_{ii}^{\alpha\beta}\cdot S_{i}^{\beta}.
\end{eqnarray}
Here, it is only left to choose which parameter, $J_{12}^{\alpha\beta}$, from the $3 \times 3$ matrix $J_{12}$ one 
wants to calculate. Without the loss of 
generality, we choose $\alpha$ to be $x$ and $\beta$ 
to be $z$. We present derivation and formula only for the $J_{12}^{xz}$ matrix element, as all other elements can be determined in the same manner. The reason for the particular choice of this matrix element lies in the fact that it is connected with the $y$ component of the DMI vector. In the paper where 4SM is introduced,\cite{E-mapping} there is additional minus sign in the formula for the mentioned $y$ component, which is correct for the antisymmetric Hamiltonian considered in that paper, however, not for a general Hamiltonian. In order to isolate $J_{12}^{xz}$, one needs to obtain the energies of four different magnetic states of the lattice as follows:
\begin{itemize}
\item state 1: $\vec{S_{1}} = (+S, 0, 0), \vec{S_{2}} = (0, 0, +S)$;\\
\item state 2: $\vec{S_{1}} = (+S, 0, 0), \vec{S_{2}} = (0, 0, -S)$;\\
\item state 3: $\vec{S_{1}} = (-S, 0, 0), \vec{S_{2}} = (0, 0, +S)$;\\
\item state 4: $\vec{S_{1}} = (-S, 0, 0), \vec{S_{2}} = (0, 0, -S)$, 
\end{itemize} and all the rest, $\vec{S}_{i \neq 1,2} = (0, +S, 0)$ or $\vec{S}_{i \neq 1,2} = (0, -S, 0)$ for all four states. The four states give four energies:
\begin{eqnarray}\label{prob8}
E_{1}&=& S\cdot J_{12}^{xz}\cdot S + \sum_{j \neq 2} S\cdot J_{1j}^{xy}\cdot S + \sum_{i \neq 1} S\cdot J_{i2}^{yz}\cdot S \nonumber
\\
&+&\sum_{i \neq 1, j \neq 2} S\cdot J_{ij}^{yy}\cdot S +S\cdot A_{11}^{xx}\cdot S + S\cdot A_{22}^{zz}\cdot S \nonumber
\\
&+&\sum_{i \neq 1,2} S\cdot A_{ii}^{yy}\cdot S,
\end{eqnarray}
\begin{eqnarray}\label{prob9}
E_{2}&=& -S\cdot J_{12}^{xz}\cdot S + \sum_{j \neq 2} S\cdot J_{1j}^{xy}\cdot S - \sum_{i \neq 1} S\cdot J_{i2}^{yz}\cdot S \nonumber
\\
&+& \sum_{i \neq 1, j \neq 2} S\cdot J_{ij}^{yy}\cdot S +S\cdot A_{11}^{xx}\cdot S + S\cdot A_{22}^{zz}\cdot S \nonumber
\\
&+& \sum_{i \neq 1,2} S\cdot A_{ii}^{yy}\cdot S,
\end{eqnarray}
\begin{eqnarray}\label{prob10}
E_{3}&=& -S\cdot J_{12}^{xz}\cdot S - \sum_{j \neq 2} S\cdot J_{1j}^{xy}\cdot S + \sum_{i \neq 1} S\cdot J_{i2}^{yz}\cdot S \nonumber
\\
&+& \sum_{i \neq 1, j \neq 2} S\cdot J_{ij}^{yy}\cdot S + S\cdot A_{11}^{xx}\cdot S + S\cdot A_{22}^{zz}\cdot S \nonumber
\\
&+& \sum_{i \neq 1,2} S\cdot A_{ii}^{yy}\cdot S,
\end{eqnarray}
and
\begin{eqnarray}\label{prob11}
E_{4}&=& S\cdot J_{12}^{xz}\cdot S - \sum_{j \neq 2} S\cdot J_{1j}^{xy}\cdot S - \sum_{i \neq 1} S\cdot J_{i2}^{yz}\cdot S \nonumber
\\
&+& \sum_{i \neq 1, j \neq 2} S\cdot J_{ij}^{yy}\cdot S + S\cdot A_{11}^{xx}\cdot S + S\cdot A_{22}^{zz}\cdot S \nonumber
\\
&+& \sum_{i \neq 1,2} S\cdot A_{ii}^{yy}\cdot S.
\end{eqnarray}
After subtracting the energies corresponding to the states 2 and 3 from the sum of energies corresponding to the states 1 and 4, and cancelling all the terms with the opposite sign and same magnitude, the only non-cancelled term is proportional to the parameter we intended to extract. This manipulation results in:
\begin{eqnarray}\label{prob12a}
&& E_{1}+E_{4}-E_{2}-E_{3}=4S^{2}\cdot J_{12}^{xz}, \nonumber
\end{eqnarray}
where it is trivial now to extract $J_{12}^{xz}$, i.e.
\begin{eqnarray}\label{prob12}
J_{12}^{xz} = \frac{E_{1}+E_{4}-E_{2}-E_{3}}{4S^{2}}.
\end{eqnarray} 
It is worth to mention that in the special case, when the off-diagonal elements are antisymmetric, i.e. $J^{\alpha\beta}=-J^{\beta\alpha}=D^{\gamma}$, where $(\alpha, \beta, \gamma)$ are $(x, y, z)$, or $(y, z, x)$, or $(z, x, y)$, equation \eqref{prob12} from this derivation reduces to equation (A5) from Ref. ~\onlinecite{E-mapping}. 

We want to point out that $\vec{S}_{1}$ and $\vec{S}_{2}$ are alternating in parallel to $x$ and $z$ directions, respectively, which corresponds to $J_{12}^{xz}$ element and $\vec{S}_{i \neq 1,2}$ being parallel or anti-parallel to $y$ direction. In case of calculation of the diagonal elements i.e. $J_{12}^{xx}$, $J_{12}^{yy}$, or $J_{12}^{zz}$, both $\vec{S}_{1}$ and $\vec{S}_{2}$ are chosen to be alternating in parallel to $x$, $y$, and $z$, respectively, and $\vec{S}_{i \neq 1,2}$ are chosen to be perpendicular to them.

\subsection{Single-ion anisotropy (A)}\label{driveA}

In order to complete the analysis, we also consider SIA parameters which govern the interaction between the spin  components of the single ion. Unlike $J_{ij}$, SIA matrix $A_{ii}$ has to be symmetric, regardless of the structural symmetry. This is the consequence of the fact that the interaction between e.g. $x$ and $z$ components is physically the same interaction as between $z$ and $x$ components on the same spin site $i$. In addition, based on the relation of $S_{i}^{2} = S^{2} = (S^{x}_{i})^{2}+(S^{y}_{i})^{2}+(S^{z}_{i})^{2}$, each component can be represented through the total spin $S$ and other two components. Due to everything stated above, in general case, for the off-diagonal elements of the SIA matrix, one needs to calculate only three upper (or lower) elements. For the diagonal element of the SIA matrix, it is sufficient to obtain two reduced terms for the diagonal part instead of all three elements separately. First of all, we present SIA Hamiltonian in explicit form:

\begin{eqnarray}\label{prob4}
H_{SIA}&=&\sum_{i} \vec{S_{i}} \cdot A_{ii} \cdot \vec{S_{i}}\nonumber \\\nonumber \\
&=&\sum_{i} \Big[ S_{i}^{x}\cdot A_{ii}^{xx}\cdot S_{i}^{x}  
+  S_{i}^{x}\cdot A_{ii}^{xy}\cdot S_{i}^{y} 
+  S_{i}^{x}\cdot A_{ii}^{xz}\cdot S_{i}^{z} \nonumber \\ \nonumber \\
&+& S_{i}^{y}\cdot A_{ii}^{yx}\cdot S_{i}^{x} 
+  S_{i}^{y}\cdot A_{ii}^{yy}\cdot S_{i}^{y}
+  S_{i}^{y}\cdot A_{ii}^{yz}\cdot S_{i}^{z} \nonumber \\ \nonumber \\
&+& S_{i}^{z}\cdot A_{ii}^{zx}\cdot S_{i}^{x}
+  S_{i}^{z}\cdot A_{ii}^{zy}\cdot S_{i}^{y} 
+  S_{i}^{z}\cdot A_{ii}^{zz}\cdot S_{i}^{z}  \Big].
\end{eqnarray}
In order to obtain the elements of the SIA matrix, one can use the same idea, as for the exchange matrix elements, with small adaptations. The first adaptation is that one needs to choose the spin-site, not the pair. The second adaptation is that one should use different procedures to obtain the off-diagonal and diagonal elements. 

\subsubsection{Off-diagonal elements of the SIA matrix}

An off-diagonal element of the SIA matrix describes the interaction between different Cartesian components of the spin at the chosen site (we take site $i=1$ without loss of generality). Namely, the elements are $A_{11}^{xy}=A_{11}^{yx}$, $A_{11}^{xz}=A_{11}^{zx}$, and $A_{11}^{yz}=A_{11}^{zy}$. The spin vector at that site should lay on the plane determined by the Cartesian components whose interaction one wants to investigate, making 45$^\circ$ angle with the two axes. All the other spins should be chosen along the complementary Cartesian axis. This means that in order to conduct 4SM using DFT calculation for 
$A_{11}^{xy}$ parameter, one should choose the four states in the following form:
\begin{itemize}
\item state 1: $\vec{S_{1}} = (+ S\sqrt{2}/2 , +S\sqrt{2}/2, 0)$;\\
\item state 2: $\vec{S_{1}} = (+S\sqrt{2}/2 , -S\sqrt{2}/2 , 0)$;\\
\item state 3: $\vec{S_{1}} = (-S\sqrt{2}/2 , +S\sqrt{2}/2 , 0)$;\\
\item state 4: $\vec{S_{1}} = (-S\sqrt{2}/2 , -S\sqrt{2}/2 , 0)$,
\end{itemize} and all the rest, $\vec{S_{i}} = (0, 0, +S)$, for all states. The four states give four energies:
\begin{eqnarray}\label{prob13}
E_{1}&=& \sum_{j>1} S\sqrt{2}/2\cdot J_{1j}^{xz}\cdot S + \sum_{j>1} S\sqrt{2}/2\cdot J_{1j}^{yz}\cdot S \nonumber
\\
&+& \sum_{i \neq 1, j>i} S\cdot J_{ij}^{zz}\cdot S + \frac{S^{2}}{2}\cdot A_{11}^{xx}+\frac{S^{2}}{2}\cdot A_{11}^{xy} \nonumber
\\
&+& \frac{S^{2}}{2}\cdot A_{11}^{yx}+\frac{S^{2}}{2}\cdot A_{11}^{yy} + \sum_{i \neq 1} S\cdot A_{ii}^{zz}\cdot S,
\end{eqnarray}
\begin{eqnarray}\label{prob14}
E_{2}&=& \sum_{j>1} S\sqrt{2}/2\cdot J_{1j}^{xz}\cdot S - \sum_{j>1} S\sqrt{2}/2\cdot J_{1j}^{yz}\cdot S \nonumber
\\
&+& \sum_{i \neq 1, j>i} S\cdot J_{ij}^{zz}\cdot S +\frac{S^{2}}{2}\cdot A_{11}^{xx}-\frac{S^{2}}{2}\cdot A_{11}^{xy} \nonumber
\\
&-& \frac{S^{2}}{2}\cdot A_{11}^{yx}+\frac{S^{2}}{2}\cdot A_{11}^{yy} + \sum_{i \neq 1} S\cdot A_{ii}^{zz}\cdot S,
\end{eqnarray}
\begin{eqnarray}\label{prob15}
E_{3}&=& - \sum_{j>1} S\sqrt{2}/2\cdot J_{1j}^{xz}\cdot S + \sum_{j>1} S\sqrt{2}/2\cdot J_{1j}^{yz}\cdot S \nonumber
\\
& +& \sum_{i \neq 1, j>i} S\cdot J_{ij}^{zz}\cdot S +\frac{S^{2}}{2}\cdot A_{11}^{xx}-\frac{S^{2}}{2}\cdot A_{11}^{xy} \nonumber
\\
&-& \frac{S^{2}}{2}\cdot A_{11}^{yx}+\frac{S^{2}}{2}\cdot A_{11}^{yy} + \sum_{i \neq 1} S\cdot A_{ii}^{zz}\cdot S,
\end{eqnarray}
and
\begin{eqnarray}\label{prob16}
E_{4}&=& - \sum_{j>1} S\sqrt{2}/2\cdot J_{1j}^{xz}\cdot S - \sum_{j>1} S\sqrt{2}/2\cdot J_{1j}^{yz}\cdot S \nonumber
\\
&+& \sum_{i \neq 1, j>i} S\cdot J_{ij}^{zz}\cdot S +\frac{S^{2}}{2}\cdot A_{11}^{xx}+\frac{S^{2}}{2}\cdot A_{11}^{xy} \nonumber
\\
&+& \frac{S^{2}}{2}\cdot A_{11}^{yx}+\frac{S^{2}}{2}\cdot A_{11}^{yy} + \sum_{i \neq 1} S\cdot A_{ii}^{zz}\cdot S.
\end{eqnarray}
After adding the energies for states 1 and 4 and subtracting from energies corresponding to states 3 and 4, the only term remaining is $A_{11}^{xy}=A_{11}^{yx}$: 
\begin{eqnarray}\label{prob17a}
E_{1}+E_{4}-E_{2}&-E_{3}=2S^{2}\cdot A_{11}^{xy}+2S^{2}\cdot A_{11}^{yx}=4S^{2}\cdot A_{11}^{xy}, \nonumber
\end{eqnarray}
hence the single-ion anisotropy parameter is obtained by formula:
\begin{eqnarray}\label{prob17}
A_{11}^{xy} = A_{11}^{yx} = \frac{E_{1}+E_{4}-E_{2}-E_{3}}{4S^{2}}.
\end{eqnarray}
The same equation (with the corresponding energies) can be written for 
all off-diagonal terms in the SIA matrix.

\subsubsection{Diagonal elements of the SIA matrix}
As we point above, in order to describe the diagonal elements of the SIA matrix, it is sufficient to calculate two reduced terms of the diagonal elements, due to the fact that $(S^{x}_{i})^{2}$ can be written as $(S^{x}_{i})^{2} =(S_{i})^{2}-(S^{y}_{i})^{2}-(S^{z}_{i})^{2}$. According to the relation, the diagonal part of the SIA Hamiltonian can be written as follows:
\begin{eqnarray}\label{probcc}
H_{SIA}^{dia}&=&S_{1}^{x} A_{11}^{xx} S_{1}^{x} + S_{1}^{y} A_{11}^{yy} S_{1}^{y}+S_{1}^{z} A_{11}^{zz} S_{1}^{z}\nonumber \\
&=&A^{xx}_{11} S_{1}^{2}+(A^{yy}_{11}-A^{xx}_{11})(S^{y}_{1})^{2} \nonumber \\
&+& (A^{zz}_{11}-A^{xx}_{11})(S^{z}_{1})^{2}.
\end{eqnarray}
The term $A^{xx}_{11} S_{1}^{2}$ is simply an additive constant. The terms $A^{yy}_{11}-A^{xx}_{11}$ and $A^{zz}_{11}-A^{xx}_{11}$ govern all information of the diagonal part of the SIA matrix. Here we present the extraction of the parameter $A^{yy}_{11}-A^{xx}_{11}$ as an example where the same strategy is valid to determine $A^{zz}_{11}-A^{xx}_{11}$ as well. The four states to be obtained are as follows:
\begin{itemize}
\item state 1: $\vec{S_{1}} = (0, +S, 0)$;\\
\item state 2: $\vec{S_{1}} = (0, -S, 0)$;\\
\item state 3: $\vec{S_{1}} = (+S, 0, 0)$;\\
\item state 4: $\vec{S_{1}} = (-S, 0, 0)$,
\end{itemize} and all the rest, $\vec{S_{i\neq1}} = (0, 0, +S)$, for all states. The energies corresponding to the chosen states are:
\begin{eqnarray}\label{prob18}
E_{1}&=& \sum_{j>1} S\cdot J_{1j}^{yz}\cdot S \nonumber \\
&+& S\cdot A_{11}^{yy}\cdot S + \sum_{i\neq 1} S\cdot A_{ii}^{zz}\cdot S,
\end{eqnarray}
\begin{eqnarray}\label{prob19}
E_{2}&=& -\sum_{j>1} S\cdot J_{1j}^{yz}\cdot S \nonumber \\
&+& S\cdot A_{11}^{yy}\cdot S + \sum_{i\neq 1} S\cdot A_{ii}^{zz}\cdot S,
\end{eqnarray}
\begin{eqnarray}\label{prob20}
E_{3}&=& \sum_{j>1} S\cdot J_{1j}^{xz}\cdot S \nonumber \\
&+& S\cdot A_{11}^{xx}\cdot S + \sum_{i\neq 1} S\cdot A_{ii}^{zz}\cdot S,
\end{eqnarray}
and
\begin{eqnarray}\label{prob21}
E_{4}&=& -\sum_{j>1} S\cdot J_{1j}^{xz}\cdot S \nonumber \\
&+& S\cdot A_{11}^{xx}\cdot S + \sum_{i\neq 1} S\cdot A_{ii}^{zz}\cdot S.
\end{eqnarray}
After summing the energies of the states 1 and 2 and subtracting those of state 3 and 4, one has the relation:
\begin{eqnarray}\label{prob22a}
E_{1}+&E_{2}-E_{3}-E_{4}=2S^{2}\cdot A_{11}^{yy}-2S^{2}\cdot A_{11}^{xx}, \nonumber
\end{eqnarray}
which results in the formula
\begin{eqnarray}\label{prob22}
A_{11}^{yy}-A_{11}^{xx} = \frac{E_{1}+E_{2}-E_{3}-E_{4}}{2S^{2}}.
\end{eqnarray}

In general material analysis, one needs to find all five elements of all the SIA matrices. Here we extend our discussion on SIA for 2D and quasi-2D materials which exhibit 3-, 4- or 6-fold rotation symmetry around out-of-plane axis. In such a case, SIA of spin-site can be described by a single parameter instead of a matrix. 

\subsubsection{Symmetry-imposed constraint in SIA of 2D and quasi-2D materials}\label{driveSYMMA}

The constraints in SIA matrix are a direct consequence of the symmetry of the crystal. In presence of 3-, 4- or 6-fold rotation symmetry around out-of-plane axis in 2D or quasi-2D material, all the elements of each matrix are equal to 0, except $A^{zz}_{11}-A^{xx}_{11}$. Here we show how the structural symmetry affects SIA of a spin-site. The symmetry can be described by the transformation matrix. Let $B$ be the transformation of the form:
\begin{eqnarray}\label{prob23}
A_{new}=B\cdot A_{old}\cdot B^{T}.
\end{eqnarray}
Moreover, if the coordinate transformation $B$ is a symmetry operation of the considered system then $A_{new}\equiv A_{old}$.  

In this particular section, we discuss coordinate transformation under rotation of the system. This means that matrix $B$ is actually the rotation matrix around $z$-axis - the axis orthogonal to layer(s) of (quasi-)2D material. We start from the general rotation matrix (any arbitrary angle - $\theta$) and later choose only four possible values compatible with the translation symmetry of the crystal systems, i.e. $\pi , \frac{2\pi}{3} , \frac{\pi}{2} , \frac{\pi}{3}$, corresponding to 2-, 3-, 4- and 6-fold rotation symmetries, respectively.

The rotation by angle $\theta$ around $z$ axis is given below in its matrix representation, as 
\begin{eqnarray}\label{eq:lnnonspbb}
R_{z}(\theta)=
\begin{bmatrix}
\cos{\theta} & \sin{\theta} & 0 \\
-\sin{\theta} & \cos{\theta} & 0 \\
0 & 0 & 1
\end{bmatrix}.
\end{eqnarray}
Now, after using $A = R_{z}(\theta)\cdot A\cdot R^{T}_{z}(\theta)$ on each matrix element, one ends up with three systems of equations. Namely, the matrix elements with indices $11$ ($22$) and $12$ ($21$) yield following system of equations:
\begin{eqnarray}\label{"*"}
\begin{cases} (A_{yy}-A_{xx})\cdot\sin^{2}{\theta}+2A_{xy}\cdot\sin{\theta}\cos{\theta}=0, \\ (A_{yy}-A_{xx})\cdot\sin{\theta}\cos{\theta}-2A_{xy}\cdot\sin^{2}{\theta}=0. \end{cases}
\end{eqnarray}
The elements $13$ ($31$) and $23$ ($32$) result in the equations:
\begin{eqnarray}\label{"**"}
\begin{cases} A_{xz}\cdot(\cos{\theta}-1)+A_{yz}\cdot\sin{\theta}=0, \\ -A_{xz}\cdot\sin{\theta}-A_{yz}\cdot(\cos{\theta}-1)=0, \end{cases}
\end{eqnarray}
while equality of the elements with indices $33$ yields trivial relation $A_{zz}=A_{zz}$.

One can easily obtain determinants of the $2$x$2$ systems \ref{"*"} and \ref{"**"}, $D^{*}$ and $D^{**}$, respectively as
\begin{eqnarray}\label{"dets"}
& D^{*} = -\sin^2{\theta}, \nonumber \\
& D^{**} = 2(1-\cos{\theta}).
\end{eqnarray}

It is well known that a homogeneous system has non-trivial solution if and only if determinant of the system is equal to 0. It is obvious that in case of 3-, 4- and 6- rotational symmetry, both determinants are different from 0, resulting in only trivial solutions for the corresponding systems of equations. This means that if 3-, 4- or 6- rotational symmetry around out-of-plane axis is present in the system, the only element of SIA matrix allowed to be non-zero is $A_{zz}-A_{xx}$.

This makes computation of the SIA part of the general spin Hamiltonian parameters five times less demanding in the case of 3-, 4- or 6-fold symmetry in 2D materials. This is due to the fact that one needs to find only one SIA matrix element, instead of five, as it is the case in the most general computation, when none of the mentioned symmetries are $a$ $priori$ present. 

\section{Example of monolayer C\lowercase{r}I$_3$}\label{example}

In this section we will present the values obtained for the exchange matrix parameters, as well as the SIA matrix parameters for pristine monolayer of CrI$_{3}$, using DFT methodology described in Appendix \ref{app}, in combination with calculations described in Section \ref{method}. We present results calculated by applying the formulae given in this study, together with the results obtained by directly applying the formulae given in the Ref. ~\onlinecite{E-mapping}. Furthermore, we comment on the differences between the formulae and potential misconceptions that could arise. 

\begin{figure}[b]
\includegraphics[width=0.98\linewidth]{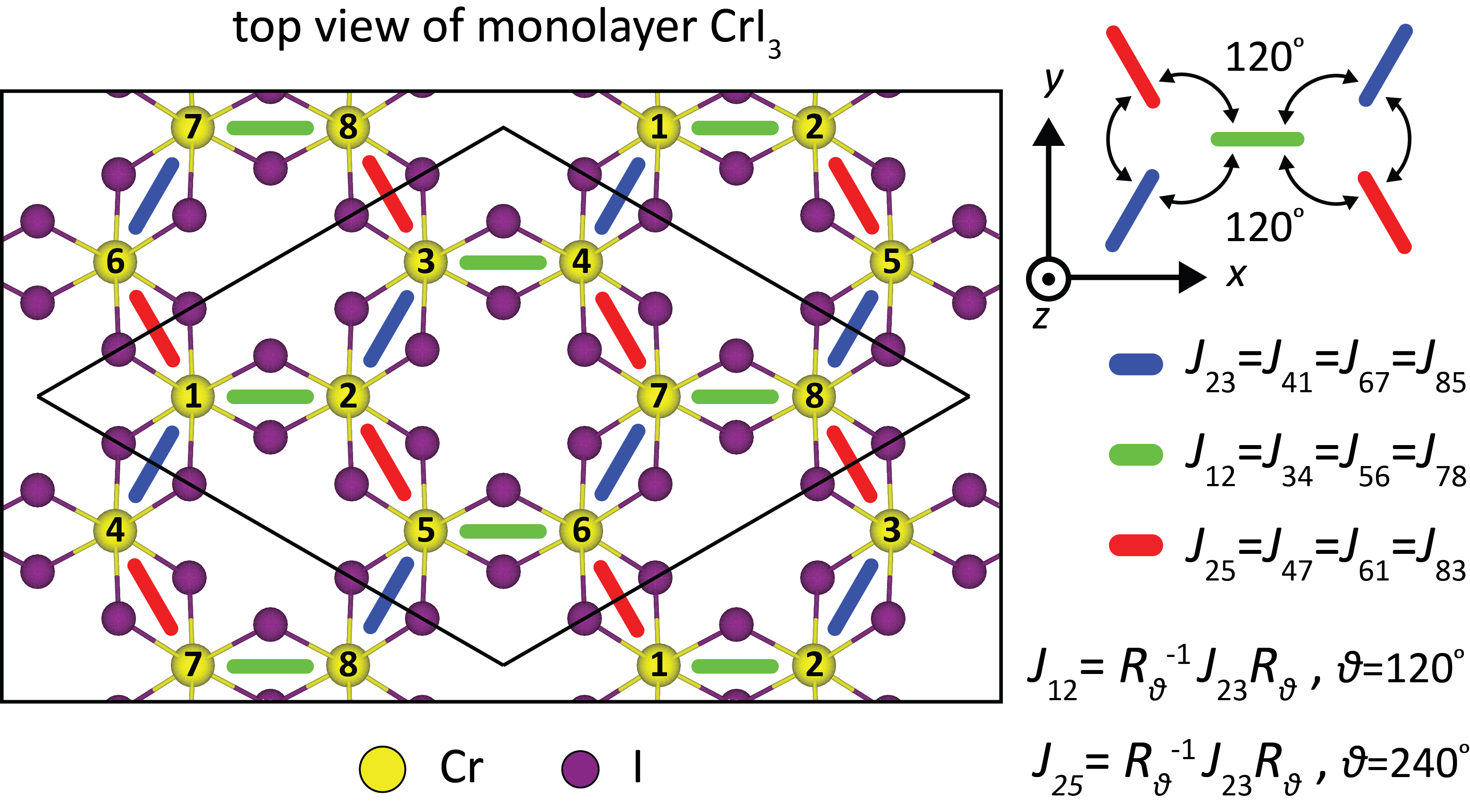}
\caption{\label{CrI32X2}
Pristine monolayer CrI$_{3}$ and its $2\times 2\times 1$ supercell (black solid line). Magnetic Cr atoms are colored yellow and labeled, while nonmagnetic I atoms are colored purple. Three Cr-Cr bonds connected with the 3-fold in-plane symmetry are shown as blue, green, and red bars with the respective magnetic exchange matrices.}
\end{figure}

We present the values of the exchange parameters between Cr atoms labeled 2 and 3 in Fig. \ref{CrI32X2}. Exchange interaction matrix between any other nearest-neighbour pair in the crystal can be calculated from the results for mentioned pair 2-3. Atom pairs 4-1, 8-5 and 6-7 are structurally identical to the pair 2-3, therefore the matrix parameters are the same as those of pair 2-3. Exchange matrices for pairs 2-5, 8-3, 4-7 and 6-1, as well as for the pairs 1-2, 3-4, 5-6 and 7-8, can be obtained from matrix characterizing the pair 2-3, by applying the 3-fold rotation (around out-of-plane axis) operations on the matrix corresponding to the pair 2-3 . Regarding the single-ion anisotropy, all atoms are described with a single matrix in case of pristine CrI$_{3}$. Moreover, the only non-zero SIA matrix element is $A_{zz}-A_{xx}$ (for details, see section \ref{driveSYMMA}).

Following the formulae given in this paper, we found all nine elements of the matrix $J_{23}$. The diagonal elements are $J_{23}^{xx} = -4.12$ meV, $J_{23}^{yy} = -4.79$ meV, $J_{23}^{zz} = -4.63$ meV, and off-diagonal elements are $J_{23}^{xy} = J_{23}^{yx} = -0.58$ meV, $J_{23}^{zx} = J_{23}^{xz} = 0.74$ meV and $J_{23}^{yz} = J_{23}^{zy} = -0.40$ meV. Obviously, matrix $J_{23}$ is symmetric. Moreover, exchange matrices of all other pairs are symmetric. This means that the antisymmetric exchange interaction, i.e. DMI, between any two nearest-neighbour spin sites is equal to 0. This is in agreement with Ref. ~\onlinecite{Moriya_DMI}, which states that if inversion center is present in the middle between two spin sites, the DM interaction between them has to be zero. The same conclusion is reached in Ref. ~\onlinecite{Xu_Kitaev} as well. 

Though the 4-state analysis in Ref. ~\onlinecite{Xu_Kitaev} was done by following the main idea of Ref. ~\onlinecite{E-mapping}, the correct formula for $J_{xy}$ parameter is given only in the supplementary material. However, no comment to generalization of the procedure from Ref. ~\onlinecite{E-mapping} is presented there, on the contrary, it is stated that detailed analysis is readily given in Ref. ~\onlinecite{E-mapping}, and in the supplementary material the main idea of the method is briefly introduced. In other words, the ambiguity on the method remained unresolved. To illustrate the impact that direct use of the formulae from Ref. ~\onlinecite{E-mapping} could have, we present below the results obtained in that manner for monolayer CrI$_{3}$. Formulae for the off-diagonal elements of exchange matrix yield $D_{23}^{z} = -0.58$ meV, $D_{23}^{y} = -0.74$ meV and $D_{23}^{x} = -0.40$ meV. This is in contradiction with the symmetry consideration by Moriya.\cite{Moriya_DMI} Furthermore, this error is even more dangerous than it may seem, since the approach gives zero average DMI experienced by each site, which is actually correct! For example, focusing on site 2, and calculating $D_{12}^{z}$ and $D_{25}^{z}$ using the formulae from Ref. ~\onlinecite{E-mapping}, one finds that together with $D_{23}^{z}$ they add up to 0, i.e. $D_{12}^{z}+D_{23}^{z}+D_{25}^{z} = 0$ (same for $x$ and $y$ components). This leads to an incorrect conclusion that DMI exists on the pair level (between two spin sites) in pristine monolayer CrI$_{3}$, but is than cancelled out by different pair-contributions. The influence of such and similar errors can progressively grow in the future with strongly increasing number of works on 2D magnetic materials. Misconceptions at the onset of a new and growing field must be avoided, and our above considerations are in function of exactly that.

Regarding the SIA matrix, since spin sites in monolayer CrI$_{3}$ exhibit 3-fold rotation symmetry around out-of-plane axis, it follows that the only non-zero SIA parameter is $A^{zz}_{11}-A^{xx}_{11}$ and it is equal to $-0.08$ meV in our case. Negative value implies that the SIA favors the out-of-plane direction instead of the in-plane one. 

\section{Summary and conclusions}\label{Conc}

To summarize, starting from the general Heisenberg spin Hamiltonian, we derived the formulae for diagonal and off-diagonal elements of 3$\times$3 matrices governing the magnetic exchange interaction between two magnetic sites ($J$) and single-ion anisotropy of a single magnetic site ($A$). The formulae are based on $ab$ $initio$ energetic calculations of the four different magnetic configurations of the magnetic crystal. The formulae as previously derived in Refs. ~\onlinecite{E-mapping, E-mapping0} are not appropriate for the general case. The generalization provided here is important and timely, due to recent realization of 2D ferromagnetism and the gaining momentum of the $ab$ $initio$ approach to magnetic systems. Otherwise, the general use of non-generalized formulae may result in wrong conclusions, e.g. about existence of DMI between the magnetic sites of the perfectly symmetric lattice of CrI$_3$. 

In order to prevent researchers from misuse of the formulae, here we presented the general and complete set of equations that is applicable to any magnetic crystal. The general formula for each element of the exchange matrıx $J$ is in the form of: 
\begin{eqnarray}\label{prob24}
J_{ij}^{\alpha \beta}= \frac{E_{1}+E_{4}-E_{2}-E_{3}}{4S^{2}},
\end{eqnarray}
where $i$ and $j$ are indices for different magnetic sites, and $\alpha$ and $\beta$ correspond to two of the three Cartesian coordinates. $E_{1-4}$ are the energies of the four states corresponding to the spins of the $i^{th}$ and $j^{th}$ site ($S_i$ and $S_j$) aligning parallel to the directions of $\pm \alpha$ and $\pm \beta$, combined with the alignment of $S_{others \neq i,j}$ with the third Cartesian axis ($\neq \alpha, \beta$).   

The formulae for the off-diagonal elements ($\alpha \neq \beta$) of $A$ are of the form:
\begin{eqnarray}\label{prob25}
A_{ii}^{\alpha \beta} = \frac{E_{1}+E_{4}-E_{2}-E_{3}}{4S^{2}},
\end{eqnarray}
where $i$ is the index of the considered single magnetic site. $\alpha$ and $\beta$ corresponds to two of the three Cartesian coordinates. $E_{1-4}$ are the energies of the four states specified as $S_{i}$ laying in $\alpha \beta$ plane and making 45$^\circ$ with $\alpha, \beta$; $-\alpha, \beta$; $\alpha, -\beta$; and $-\alpha, -\beta$. The spins of other sites, $S_{other \neq i}$ point along the third Cartesian axis. 
For the diagonal elements ($\alpha = \beta$) one needs to calculate their reduced form: 
\begin{eqnarray}\label{prob26}
A_{ii}^{\alpha \alpha} - A_{ii}^{xx} = \frac{E_{1}+E_{2}-E_{3}-E_{4}}{2S^{2}},
\end{eqnarray}
where $\alpha$ is either $y$ or $z$. $E_{1-4}$ are the energies of the four states of $S_i$ pointing along $\pm \alpha$ and $\pm x$, with $S_{other\neq i}$ orthogonal to both $\alpha$ and $x$. 
In addition to the general assessment of SIA, we present SIA analysis for specifically 2D-like crystals which exhibit 3-, 4- or 6- fold rotation symmetry. In such cases, there are three possible results: (i) the in-plane anisotropy where $A_{zz}-A_{xx} > 0$; (ii) out-of-plane anisotropy where $A_{zz}-A_{xx} < 0$; (iii) no anisotropy $A_{zz}-A_{xx} = 0$. In the last case, SIA is the 0-matrix, i.e. $A_{xx} = A_{yy} = A_{zz} = A$. In terms of total Heisenberg Hamiltonian, SIA is just an additive constant equal to $AS^{2}$ and does not influence the Hamiltonian spectrum.

Finally, the formulae derived here were applied to the pristine crystal of monolayer CrI$_{3}$. The results are given together with those obtained by strictly following Refs. ~\onlinecite{E-mapping, E-mapping0} in order to illustrate the error one could make if not carefully applying four-state methodology on the system of interest. For example, the general formula presented here results in no DMI between adjacent Cr atoms of monolayer CrI$_{3}$, which is consistent with Moriya's theorem\cite{Moriya_DMI}, while the direct application of previously established formulae yields finite DMI as an erroneous result.

By clearly stating the complete formalism, we hope to remove any potential doubt or eventual error that might have arisen in the field. Such error is rather likely when estimating the off-diagonal exchange parameters, which correspond to the DMI. In fact the DMI in 2D materials is an entirely new subject, hence it is timely and important to clarify and generalize the relevant formulae, and thereby avoid erroneous results and conclusions at the onset of the exciting field of research.

\begin{acknowledgments}
This work was supported by the Research Foundation-Flanders (FWO-Vlaanderen) and the Special Research Funds of the University of Antwerp (TOPBOF). The computational resources and services used in this work were provided by the VSC (Flemish Supercomputer Center), funded by Research Foundation-Flanders (FWO) and the Flemish Government -- department EWI.
\end{acknowledgments}

\appendix*
\section{Computational DFT methodology}\label{app}
For the purpose of applying the formulae derived in this paper on a specific material, we performed DFT-based calculations using Vienna $Ab$ $initio$ Simulation Package (VASP)\cite{vasp1,vasp2} within the projector augmented wave (PAW)\cite{Blochl} method. The electron exchange and correlation is described as the Perdew-Burke-Ernzerhof (PBE)\cite{perdew} form of the generalized gradient approximation (GGA). 3d$^{5}$ and 4s$^{1}$ electrons of Cr atom and 5s$^{2}$ and 5p$^{5}$ electrons of I atom were considered as valance electrons. To construct a 2D crystal structure, we set vacuum height of 15 \AA{}. In order to isolate the spin (pair) site from the neighboring unit cell, $2\times 2\times 1$ supercell was considered as the unit cell for the Four-state calculation and $3\times 3\times 1$ $k$-point sampling was chosen. Cut-off energy for a plane wave basis set was chosen as 500 eV. Energy convergence criterion was set to 10$^{-5}$ eV between two successive iterations. On-site Coulomb repulsion parameter, $U$ was taken as 4 eV for magnetic Cr atoms and 0 for I atoms.\cite{Dudarev} For the Brillouin zone integration we used Gaussian smearing of 0.01 meV.


\begin{thebibliography}{99}

\bibitem{Huang_2Dmag} B. Huang \textit{et al.}, Nature \textbf{546}, 270 (2017).

\bibitem{Gibertini_magn2D} M.Gibertini, M. Koperski, A. F. Morpurgo, K. S. Novoselov, Nat. Nanotechnol. \textbf{14}, 408 (2019).

\bibitem{Sivadas} N. Sivadas, M. W. Daniels, R. H. Swendsen, S. Okamoto, D. Xiao,  Phys. Rev. B \textbf{91}, 235425 (2015).

\bibitem{Lado} J. L. Lado, J. Fernández-Rossier, 2D Mater. \textbf{4}, 3 (2017).

\bibitem{Xu_Kitaev} C. Xu, J. Feng, H. Xiang, L. Bellaiche, npj Comput. Mater. \textbf{4}, 57 (2018).

\bibitem{Deng_gatetune} Y. Deng, Nature \textbf{563}, 94 (2018).

\bibitem{Gong} C. Gong, L. Li, Z. Li \textit{et al.}, Nature \textbf{546}, 265 (2017).

\bibitem{Bonilla_VSe} M. Bonilla \textit{et al.}, Nat. Nanotechnol. \textbf{13}, 289 (2018).

\bibitem{MWtheorem} N. D. Mermin and H. Wagner, Phys. Rev. Lett. \textbf{17}, 1133 (1966).

\bibitem{Moriya_DMI} T. Moriya, Phys. Rev. \textbf{120}, 91 (1960).

\bibitem{Dzia_DMI} I. E. Dzialoshinskii, Sov. Phys. JETP \textbf{5}, 1259 (1957).

\bibitem{Spinice} C. Castelnovo, R. Moessner, S. L. Sondhi,  Nature \textbf{451}, 42 (2008).

\bibitem{Skyrmions} A. Fert, V. Cros, and J. Sampaio, Nat. Nanotechnol. \textbf{8}, 152 (2013).

\bibitem{Yang} H. Yang, G. Chen, A. A. C. Cotta \textit{et al.},  Nature Mater \textbf{17}, 605 (2018). 

\bibitem{E_DMI} J. Liu, M. Shi, and J. Lu, Phys. Rev. B \textbf{97}, 8 (2018).

\bibitem{Magnon} A. Bergman \textit{et al.}, Phys. Rev. B \textbf{81}, 1 (2010).

\bibitem{Spinwave} K. Zakeri \textit{et al.}, Phys. Rev. Lett. \textbf{104}, 1 (2010).

\bibitem{Spintronics} R. A. Duine, K. J. Lee, S. S. P. Parkin, and M. D. Stiles, Nature Physics \textbf{14}, 217 (2018).

\bibitem{Chittari} B. L. Chittari, Y. Park, L. Dongkyu, M. Han, A. H. MacDonald, E. Hwang, and J. Jung, Phys. Rev. B \textbf{94}, 184428 (2016).

\bibitem{E-mapping0} H. J. Xiang, E. J. Kan, S. H. Wei, M. H. Whangbo, and X. G. Gong, Phys. Rev. B \textbf{84}, 224429 (2011).

\bibitem{E-mapping} H. Xiang, C. Lee, H. J. Koo, X. Gong, and M. H. Whangbo, Dalt. Trans. \textbf{42}, 823 (2013).

\bibitem{gen_notgenH} J. Liu, H. J. Koo, H. Xiang, R. K. Kremer, M. H. Whangbo, J. Chem. Phys. \textbf{141}, 124113 (2014).

\bibitem{gen_notgenH1} P. S. Wang, W. Ren, L. Bellaiche, and H. J. Xiang, Phys. Rev. Lett. \textbf{114}, 147204 (2015).

\bibitem{SE1} C. Xu, B. Xu, B. Dupe, and L. Bellaiche, Phys. Rev. B \textbf{99}, 104420 (2019). 

\bibitem{SE2} K. Riedl, D. Guterding, H. O. Jeschke, M. J. P. Gingras, and R. Valentí, Phys. Rev. B \textbf{94}, 014410 (2016).

\bibitem{vasp1} G Kresse and J. Hafner, Phys. Rev. B \textbf{47}, 558(R) (1993).

\bibitem{vasp2} G. Kresse and J. Furthmuller, Phys. Rev. B \textbf{54}, 11169 (1996).

\bibitem{perdew} J. P. Perdew, K. Burke, and M. Ernzerhof, Phys. Rev. Lett.  \textbf{77}, 3865 (1996).

\bibitem{Blochl} P. E. Blochl, Phys. Rev. B \textbf{50}, 17953 (1994).

\bibitem{Dudarev} S. L. Dudarev, G. A. Botton, S. Y. Savrasov, C. J. Humphreys, and A. S. Sutton, Phys. Rev. B  \textbf{57}, 1505 (1998). 

\end{thebibliography}
\end{document}